\documentclass{article}

\usepackage{spconf,amsmath,amsfonts,amssymb,graphicx}

\DeclareMathOperator*{\minimize}{\text{minimize}}

\usepackage{adjustbox} 

\usepackage{amsmath,amsfonts,bm}

\def\eqref#1{(\ref{#1})}

\def\1{\bm{1}}

\DeclareMathAlphabet{\mathsfit}{\encodingdefault}{\sfdefault}{m}{sl}
\SetMathAlphabet{\mathsfit}{bold}{\encodingdefault}{\sfdefault}{bx}{n}

\DeclareMathOperator*{\argmin}{arg\,min}

\usepackage{booktabs}
\usepackage{wrapfig}
\usepackage{array}
\usepackage{tablefootnote}
\usepackage{subcaption}
\usepackage{pifont}
\usepackage{colortbl}
\usepackage{hyperref} 
\usepackage{bm}

\DeclareMathAlphabet\mathbfcal{OMS}{cmsy}{b}{n}

\usepackage{color}

\newcommand{\modl}{\textsc{MoDL}}
\newcommand{\us}{\textsc{SMUG}}
\newcommand{\usold}{\textsc{SMUGv0}}
\usepackage{cite}
\usepackage{tcolorbox}

\title{SMUG: Towards Robust MRI Reconstruction by Smoothed Unrolling}
\name{Hui Li$^1$ \, Jinghan Jia$^2$ \, Shijun Liang$^2$ \, Yuguang Yao$^2$ 
\, Saiprasad Ravishankar$^{2}$ \,  Sijia Liu$^2$}
\address{$^1$Huazhong University of Science and Technology, China
\\
$^2$Michigan State University, East Lansing, MI, USA 
\thanks{$^1$The work is done during remote internship at MSU.}
}
\begin{document}
\ninept
\maketitle
%



\newcommand{\Def}[0]{\mathrel{\mathop:}=}

\newcommand{\Blue}[1]{\textcolor[rgb]{0.00,0.00,1.00}{#1}}
\newcommand{\tiger}[1]{\Blue{#1}}
 
\newcommand{\bx}{\mathbf{x}}
\newcommand{\by}{\mathbf{y}}
\newcommand{\bz}{\mathbf{z}}
\newcommand{\bX}{\mathbf{X}}
\newcommand{\bh}{\mathbf{h}}
\newcommand{\bu}{\mathbf{u}}
\newcommand{\bg}{\mathbf{g}}
\newcommand{\din}{\mathcal D^{\texttt{tr}}}
\newcommand{\dint}{\tilde{\mathcal D}^{\texttt{tr}}}
\newcommand{\dout}{\mathcal D^{\texttt{val}}}
\newcommand{\doutt}{\tilde{\mathcal D}^{\texttt{val}}}
\newcommand{\btheta}{\boldsymbol{\theta}}
\newcommand{\bphi}{\boldsymbol{\phi}}
\newcommand{\bdelta}{\boldsymbol{\delta}}
\newcommand{\grad}{\nabla}
\newcommand{\egrad}{\widehat{\nabla}}
\newcommand{\task}{\mathcal{T}}
\newcommand{\rnn}{\text{{\tt RNN}}\xspace}
\newcommand{\EI}{\text{EI}}
\newcommand{\lopt}{\text{LO}}
\newcommand{\V}{\mathbb{V}}
\newcommand{\indep}{\perp \!\!\! \perp}
\newcommand{\pphi}[1]{\frac{\partial #1}{\partial \bphi}}

\newcommand{\pre}{\textsc{p}}
\newcommand{\ft}{\textsc{f}}
\newcommand{\adv}{\textsc{a}}
\newcommand{\sta}{\textsc{n}}
\newcommand{\super}{\textsc{s}}
\newcommand{\self}{\textsc{ss}}
\newcommand{\Sp}{\textit{supervised}}
\newcommand{\Ss}{\textit{self-supervised}}

\newcommand{\cL}{\mathcal{L}}
\newcommand{\cD}{\mathcal{D}}

\begin{abstract}
Although deep learning ({DL}) has gained much popularity for accelerated magnetic resonance imaging ({MRI}), recent studies have shown that DL-based MRI reconstruction models could be over-sensitive to  tiny input perturbations (that are called `adversarial perturbations'), which cause unstable, low-quality reconstructed images.
This raises the question of how to design robust DL methods for  MRI reconstruction. 
To address this problem, we propose  a novel image reconstruction framework, termed \textsc{\underline{Sm}oothed \underline{U}nrollin\underline{g}} ({\us}), which advances a deep unrolling-based MRI reconstruction model using a randomized smoothing (RS)-based robust learning operation. RS, which improves the tolerance of a model against input noises, has been widely used in the design of adversarial defense for image classification. Yet, we find that  the conventional design that applies RS to  the entire DL process is ineffective for  MRI reconstruction. We show that {\us} addresses the above issue by customizing the RS operation based on the unrolling  architecture  of the DL-based MRI reconstruction model. Compared to   the vanilla RS approach and several variants of {\us},  we show that 
{\us} improves the robustness of MRI reconstruction with respect to a diverse set of  perturbation sources, including perturbations to input measurements, different measurement sampling rates, and different unrolling steps. Code for {\us} will be available at \texttt{\url{https://github.com/LGM70/SMUG}}.

\end{abstract}
\section{Introduction}
\label{sec:intro}




Magnetic resonance imaging (MRI) is a widely used imaging modality in clinical practice that is used to image both anatomical structures and physiological functions. However, the data collection in MRI is sequential and slow. Thus, many methods\cite{lustig2008compressed,yang2010fast,Aggarwal2019MoDL:Problems} have been developed to provide accurate image reconstructions from limited (rapidly collected) data.

Recently, deep learning (DL) has become a powerful tool   to solve  image reconstruction and inverse problems in general ~\cite{Schlemper2019Sigma-net:Reconstruction,Ravishankar2018DeepReconstruction,Aggarwal2019MoDL:Problems,Schlemper2018AReconstruction}.  In this paper, we focus on the application of DL to MRI reconstruction.
Among DL-based methods, image or sensor domain denoising networks are well-known. The most prevalent deep neural networks include the U-Net~\cite{Unet} and variants~\cite{han2018framing, lee2018deep} that are adapted to correct the artifacts in MRI reconstructions from undersampled data. 
Hybrid-domain methods that combine neural networks together with imaging physics such as forward models have become quite popular. 
One such state-of-the-art algorithm is the unrolled network scheme, MoDL~\cite{Aggarwal2019MoDL:Problems} that mimics an iterative algorithm to solve the regularized inverse problem in MRI reconstruction. Its variants have achieved top performance in recent open data-driven competitions. 
 
However, many studies \cite{antun2020instabilities,zhang2021instabilities, gilton2021deep} have demonstrated that DL-based MRI reconstruction models suffer from a lack of robustness. It has been shown that DL-based models are vulnerable to tiny input perturbations \cite{antun2020instabilities, zhang2021instabilities}, changes in measurement sampling rate \cite{antun2020instabilities}, and changes in the number of iterations of the model~\cite{gilton2021deep}. In these scenarios, the reconstructed images generated by DL-based models are of poor quality, which may lead to false diagnoses and adverse clinical consequences.

Although many defense methods~\cite{madry2017towards,zhang2019theoretically,cohen2019certified,salman2020denoised} were proposed to address the lack of robustness of DL models on the image classification task, the approaches of robustifying DL-based MRI reconstruction models are under-developed due to their regression-based learning objectives.
Randomized smoothing (RS) 
and its variants~\cite{cohen2019certified, salman2020denoised,zhang2022robustify} are quite popular adversarial defense methods in image classification. Different from conventional defense methods\cite{madry2017towards,zhang2019theoretically} which generate empirical robustness and are prone to fail against stronger attacks, RS  guarantees the model's robustness within a small sphere around the input image \cite{cohen2019certified}, which is vital for medical applications like MRI. A recent preliminary work attempted to apply RS to DL-based MRI reconstruction in an end-to-end (E2E) manner~\cite{wolfmaking}.

Given the advantages of RS and deep unrolling-based (hybrid domain) image reconstructors, we propose a novel approach dubbed \textsc{\underline{Sm}oothed \underline{U}nrollin\underline{g}} ({\us}) to mitigate the lack of robustness of DL-based MRI reconstruction models by systematically integrating RS into {\modl}~\cite{Aggarwal2019MoDL:Problems} architectures.
Instead of inefficient conventional \ref{eq: denoised smoothing mri} \cite{wolfmaking}, we apply  RS  in every unrolling step and on intermediate unrolled denoisers in {\modl}. We follow the `pre-training + fine-tuning' technique \cite{zoph2020rethinking,salman2020denoised}, adopting a mean square error (MSE) loss for pre-training and proposing an unrolling stability (UStab) loss along with the vanilla {\modl} reconstruction loss for fine-tuning.
Different from the existing art, our \textbf{contributions} are summarized as follows.\\
\noindent $\bullet$ \ We propose {\us} that systematically integrates RS with {\modl} using an deep unrolled architecture.\\
\noindent $\bullet$ \  We study in detail where to apply RS in the unrolled architecture for better performance and propose a novel unrolling loss to improve training efficiency.\\
\noindent $\bullet$ We compared our methods with two related baselines: vanilla {\modl}~\cite{Aggarwal2019MoDL:Problems} and \ref{eq: denoised smoothing mri} \cite{wolfmaking}. Extensive experiments demonstrate the significant effectiveness of our proposed method on the major types of instabilities of {\modl}.

\section{Preliminaries and Problem Statement}
\label{sec: preliminaries}

In this section, we provide a brief background on MRI reconstruction and motivate the problem of our interest.

\vspace*{1mm}
\noindent \textbf{Setup of MRI reconstruction.} MRI reconstruction is an 
 ill-posed inverse problem~\cite{compress}, which aims to reconstruct the original signal $\mathbf x \in\mathbb{C}^{q}$ from its  measurement $\mathbf y \in  \mathbb{C}^p$ with $p < q$. The imaging system in MRI can be modeled as a linear system $\mathbf y \approx  \mathbf A \mathbf x $, {where $\mathbf A$ may take on different forms for single-coil or parallel (multi-coil) MRI, etc.}
 For example,  $\mathbf A = \mathbf S \mathbf F$ in the single-channel Cartesian MRI acquisition setting, where $\mathbf F$ is the 2-D discrete Fourier
transform and $\mathbf S$ is a (fat) Fourier subsampling matrix, and its sparsity is controlled by the measurement sampling rate or acceleration factor.
 With the linear observation model, 
 MRI reconstruction is often formulated as
\begin{align}
    \hat{\bx}=\underset{\bx}{\arg\min} ~ \|\mathbf A \bx - \by \|^{2}_2 + \lambda \mathcal{R}(\bx),
    \label{eq:inv_pro}
\end{align}
where   $\mathcal{R}(\cdot)$ is a regularization function (\textit{e.g.}, $\ell_1$ norm to impose a sparsity prior), and $\lambda > 0$ is a regularization parameter.  





\underline{Mo}del-based reconstruction using \underline{D}eep \underline{L}earned priors
({\modl}) \cite{Aggarwal2019MoDL:Problems} was proposed recently as a deep learning-based alternative approach to solving Problem~\eqref{eq:inv_pro}, and
has attracted much interest as it 
merges the power of model-based reconstruction schemes with DL.
In {\modl}, the hand-crafted regularizer $\mathcal R$ is replaced by a learned network-based prior (involving a deep convolutional neural network (CNN)).
The corresponding formulation is
{
\begin{align}
    \hat{\bx}_{\boldsymbol \theta}=\underset{\bx}{\arg\min} ~  \|\mathbf A \bx - \by \|^{2}_2 + \lambda \| \bx - \cD_{\boldsymbol \theta} (\bx) \|_2^2,
\label{eq:altmin}    
\end{align}
}%
where $\cD_{\boldsymbol \theta} (\bx)$  denotes a 
deep network with  parameters $\boldsymbol \theta$, with input $\mathbf x$.
To obtain $  \hat{\bx}_{\boldsymbol \theta}$,   an alternating process based on variable splitting is typically used, which involves the following two steps \ding{172}--\ding{173}, executed iteratively. 

\noindent 
\ding{172} \textbf{Denoising step:} Given an updated solution $\mathbf x_n$ at the $n$th iteration (also known as `unrolling step'), {\modl} 
uses the ``denoising" network to obtain $\mathbf z_n :=\cD_{\boldsymbol \theta} (\bx_n)$. 

\noindent 
\ding{173}   \textbf{Data-consistency (DC) step:}  {\modl} then solves a least-squares problem with a denoising prior as
$
\mathbf x_{n+1} = \argmin_{\mathbf x} 
  \|\mathbf A \bx - \by \|^{2}_2 + \lambda \| \bx - \mathbf z_n \|_2^2
$, which is convex (fixed $\mathbf z_n$) with closed-form solution. 
 
In {\modl}, this alternating process is unrolled for a few iterations and the denoising network's weights are trained end-to-end in a supervised manner. 
For the rest of this paper, the function $\mathbf x_{\modl}(\cdot)$ denotes the image reconstruction process of {\modl}. 



\vspace*{1mm}
\noindent \textbf{Motivation: Lack of robustness in {\modl}.}
It was shown in \cite{antun2020instabilities} that DL may lack stability in image reconstruction, especially when facing tiny, almost undetectable input perturbations. 
Such perturbations are known as `adversarial attacks', and have been well studied in  DL for image classification~\cite{Goodfellow2015explaining}. 
Let $\boldsymbol \delta$ denote a small   perturbation of a point that falls in an $\ell_\infty$ ball of radius $\epsilon$, \textit{i.e.}, $\| \boldsymbol \delta \|_\infty \leq \epsilon$. Adversarial attack then corresponds to the worst-case input perturbation $\boldsymbol \delta$ that maximizes the reconstruction error, \textit{i.e.}, 
\begin{equation}
\begin{array}{ll}
\displaystyle \minimize_{\| \bdelta \|_\infty \leq \epsilon }     &  - \|  
{\bx}_\text{{\modl}} (\mathbf A^H \mathbf y + \boldsymbol \delta) -  \mathbf t \|_2^2,
\end{array}
  \label{eq: atk_perturb}
\end{equation}
where $\mathbf t$ is a target  image (\textit{i.e.}, label), the operator $\mathbf A^H $  transforms the measurements $\mathbf y$ to the image domain, and $\mathbf A^H \mathbf y$ is the input (aliased) signal for the {\modl}-based   reconstruction network.  Given a   {\modl} model,  problem \eqref{eq: atk_perturb} can be effectively solved using the iterative projected gradient descent (PGD) method~\cite{madry2017towards}. The resulting solution is   called `PGD attack'.

\begin{figure}[htb]
    
\begin{tabular}[b]{cccc}
        \includegraphics[width=.2\linewidth, trim=70 10 70 10]{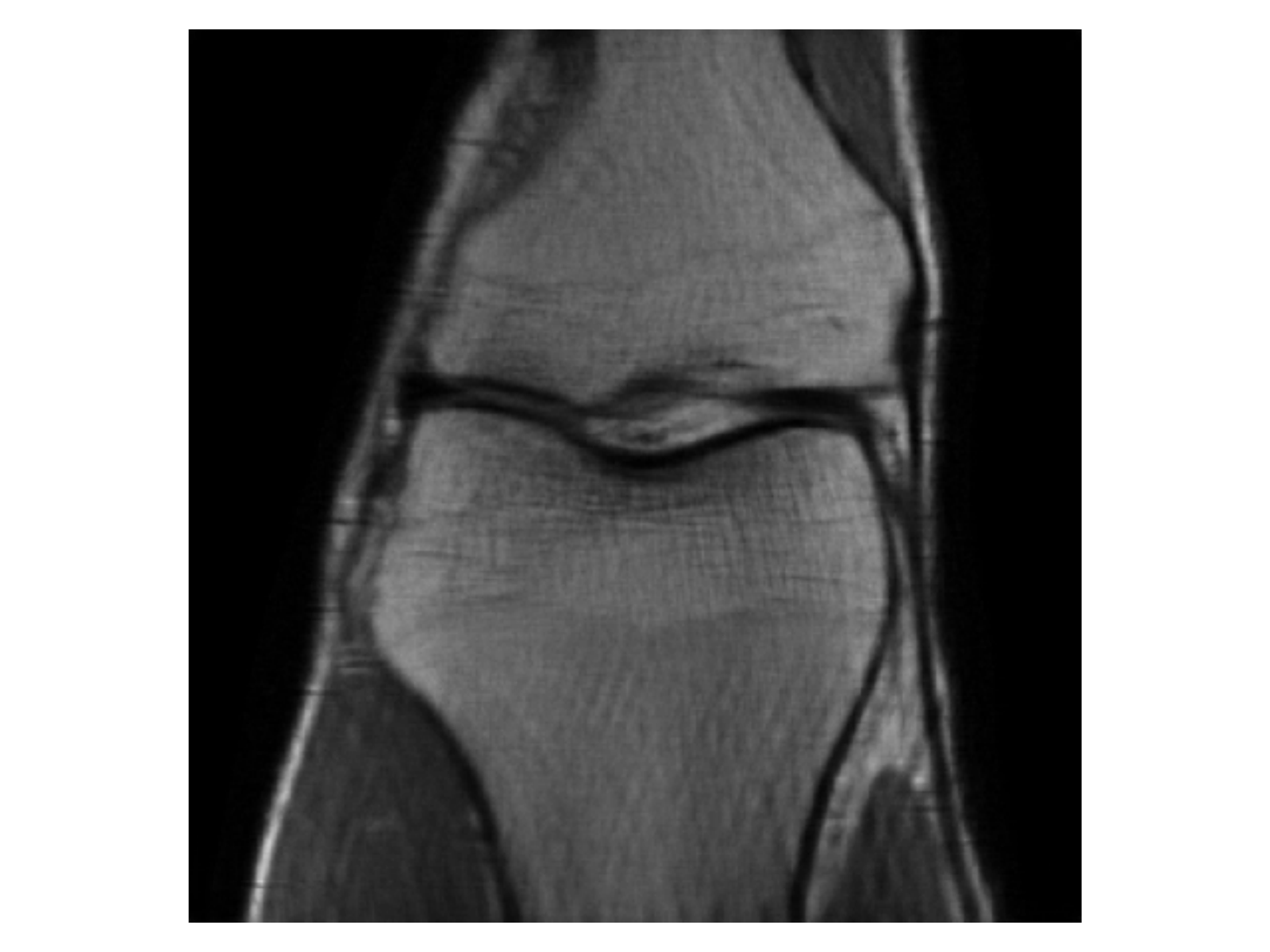}
        &
        \includegraphics[width=.2\linewidth, trim=70 10 70 10]{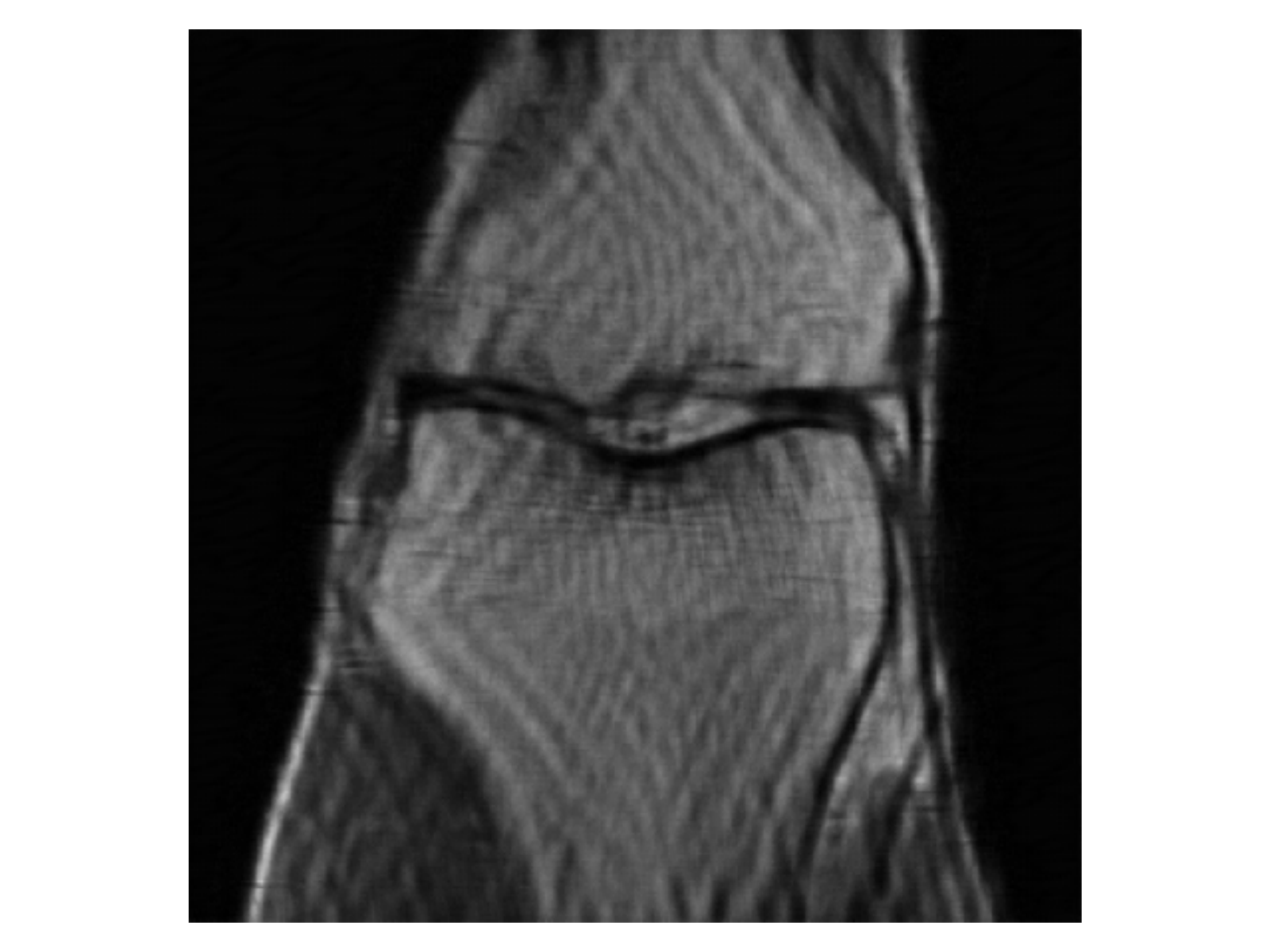}
        &
        \includegraphics[width=.2\linewidth, trim=70 10 70 10]{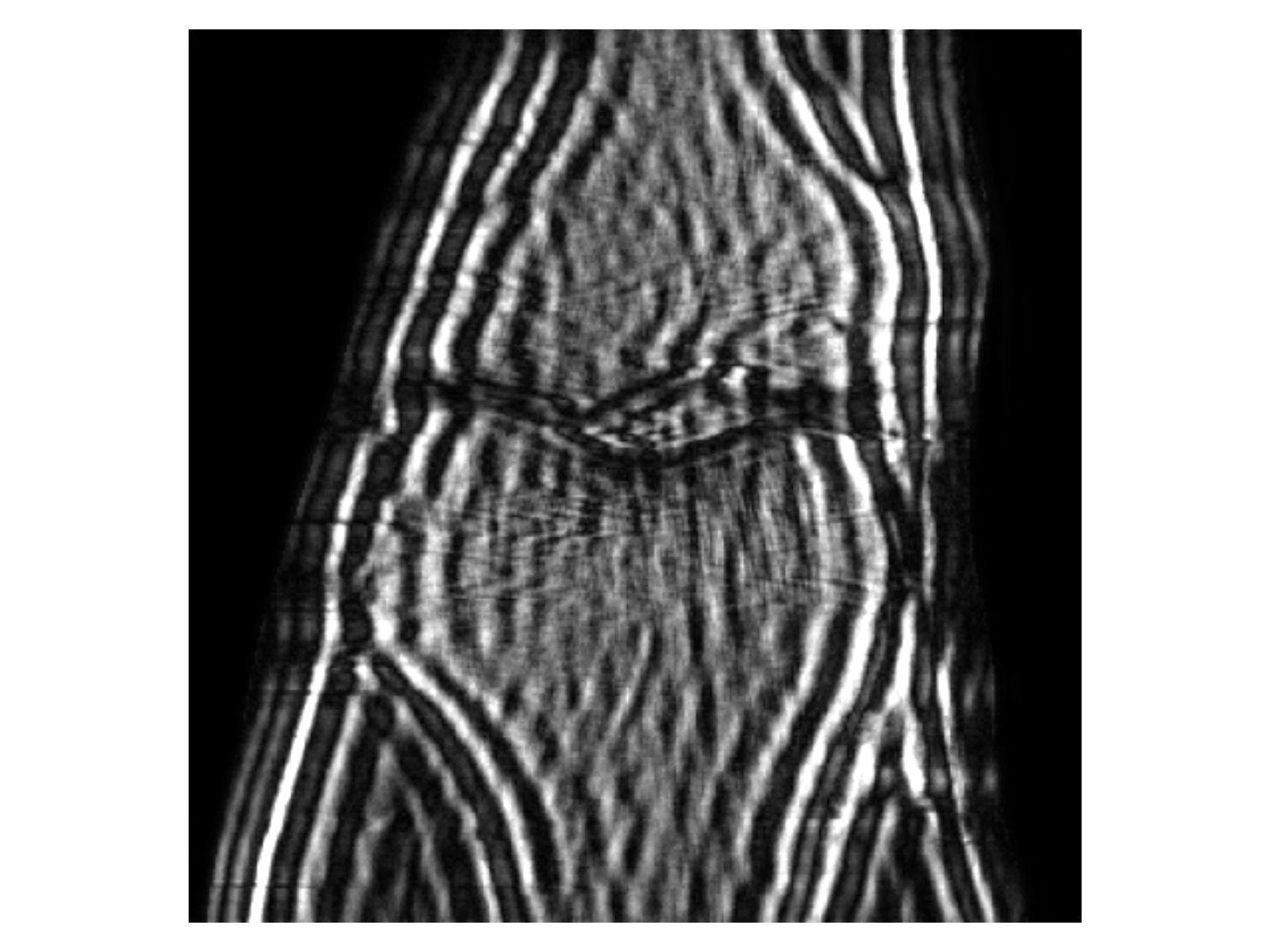}
        &
        \includegraphics[width=.2\linewidth, trim=70 10 70 10]{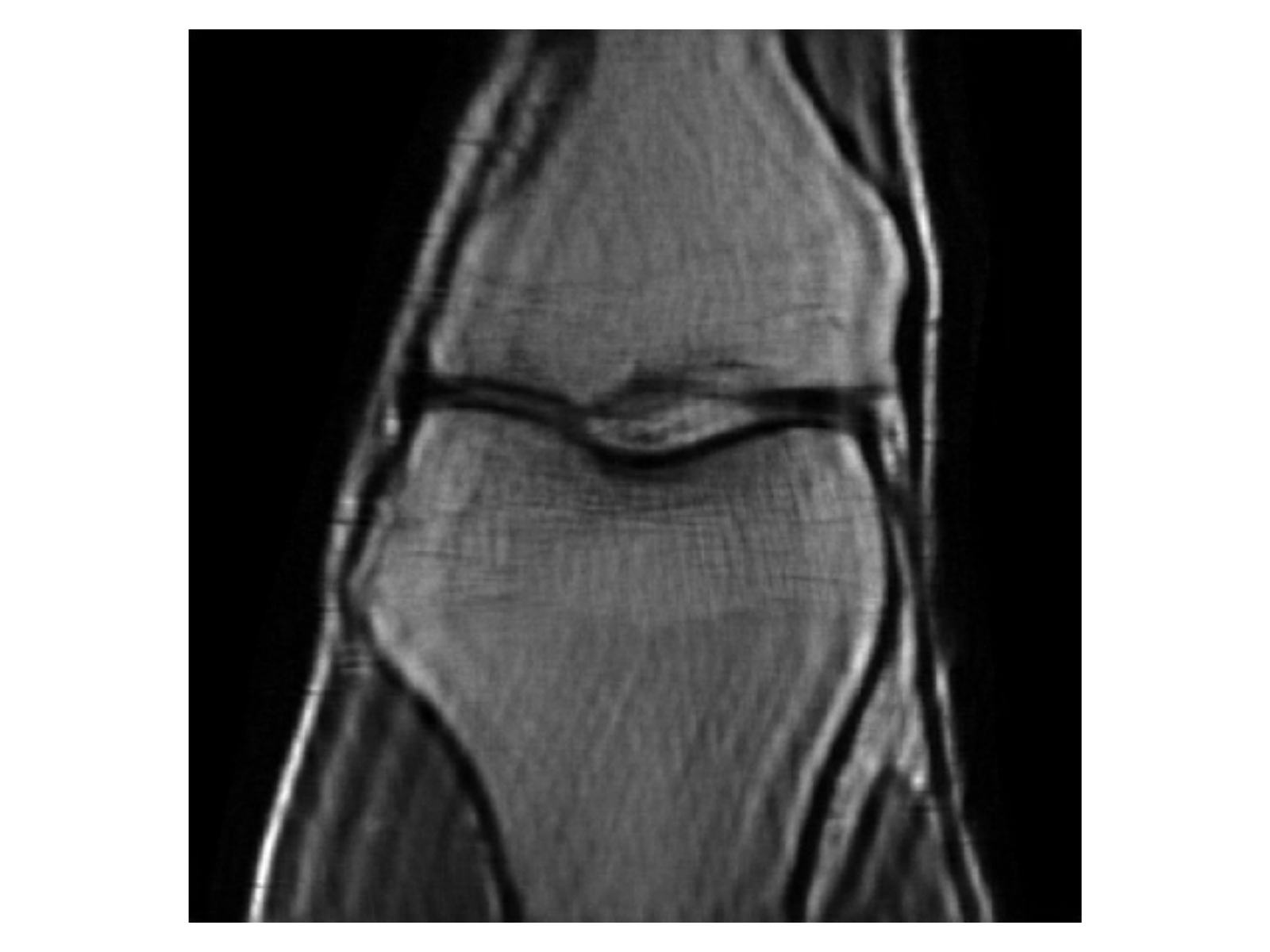}
        \\[-0pt]
        \scriptsize{(a)} 
        & 
        \scriptsize{(b)} 
        &  
        \scriptsize{(c)} 
        &
        \scriptsize{(d)}

\end{tabular}
\vspace*{-5mm}
\caption{\footnotesize{
{\modl}'s instabilities against perturbations to input data, the measurement sampling rate, and the number of unrolling steps used at testing time shown on an image from the \texttt{fastMRI} \cite{zbontar2018fastmri}
dataset. We refer readers to  Sec.\,\ref{sec: experiment} for more experiment details.
(a) {\modl} reconstruction from benign (\textit{i.e.}, clean) measurement with $ 4\times$ acceleration (\textit{i.e.}, 25\% sampling rate) and 8 unrolling steps. (b) {\modl} reconstruction from adversarial input of perturbation strength $\epsilon = 0.002$ (other settings are same as
 (a)). 
(c) {\modl} reconstruction  from clean measurement   with $ 2\times$ acceleration (\textit{i.e.}, 50\% sampling rate) and  using 8 unrolling steps.
(d) {\modl} reconstruction from clean measurement with $ 4\times$ acceleration and using  16 unrolling steps.
}}
\label{fig: weakness}
\vspace*{-4mm}
\end{figure}

In \textbf{Fig.\,\ref{fig: weakness}-(a)} and \textbf{(b)}, we demonstrate an example of the reconstructed image 
$\bx_\text{{\modl}}$ 
from a benign input (\textit{i.e.}, clean and unperturbed input) and a PGD attacked input, respectively. As we can see, the quality of reconstructed image significantly degrades in the presence of adversarial (very small) input perturbations. 
Although robustness against adversarial attacks is a primary focus of this work, \textbf{Fig.\,\ref{fig: weakness}-(c)} and \textbf{(d)} show two other types of instabilities that {\modl} may suffer at testing time: the change of the measurement sampling rate (which leads to `perturbations' to the sparsity of sampling mask in $\mathbf A$) \cite{antun2020instabilities}, and the number of unrolling steps \cite{gilton2021deep} used in {\modl} for test-time image reconstruction.
We observe that a over sampling mask  (\textbf{Fig.\,\ref{fig: weakness}-(c)}) and a larger number of unrolling steps (\textbf{Fig.\,\ref{fig: weakness}-(d)}), which deviate from the training-time setting of  {\modl}, can lead to much poorer image reconstruction performance than the original setup (\textbf{Fig.\,\ref{fig: weakness}-(a)}) even in the absence of an adversarial input.
In Sec.\,\ref{sec: experiment}, we will show that our proposed approach (originally designed for improving {\modl}'s robustness against adversarial attacks) 
yields resilient reconstruction performance against all perturbation types shown in \textbf{Fig.\,\ref{fig: weakness}}.
 
\vspace*{1mm}
\noindent \textbf{Randomized smoothing (RS).} 
RS creates multiple random noisy copies of input data and takes an averaged output over these noisy inputs so as to gain  robustness against input noises \cite{cohen2019certified}. Formally, given a base function $f(\mathbf x)$, RS turns this base function to a smoothing version  $g(\mathbf x) \Def \mathbb{E}_{\boldsymbol \nu \sim\mathcal{N}(\mathbf 0, \sigma^2\mathbf I)} [ f( \bx + \boldsymbol \nu) ]$, where $\boldsymbol \nu \sim\mathcal{N}(\mathbf 0, \sigma^2\mathbf I)$ denotes the Gaussian distribution with zero mean and $\sigma^2$-valued variance. 
In the literature, RS  has been used as an effective adversarial defense in image classification \cite{cohen2019certified,salman2020denoised,zhang2022how}.
However, it remains elusive whether or not RS is an effective solution to improving robustness of {\modl} and other image reconstructors. A preliminary study towards this direction was provided by \cite{wolfmaking}, which integrates RS with image reconstruction in an end-to-end (\textbf{E2E}) manner.   For {\modl}, this yields
\begin{align}
    g(\mathbf A^H \mathbf y)=\mathbb{E}_{\boldsymbol \nu \sim\mathcal{N}(\mathbf 0, \sigma^2\mathbf I)} [ 
    {\mathbf{x}}_{\text{\modl}} (\mathbf A^H \mathbf y + \boldsymbol \nu) ]
    .
    \label{eq: denoised smoothing mri}
    \tag{RS-E2E}
\end{align}

\begin{figure}[htb]
\vspace*{-3mm}
    \centering
        \includegraphics[width=0.46\textwidth]{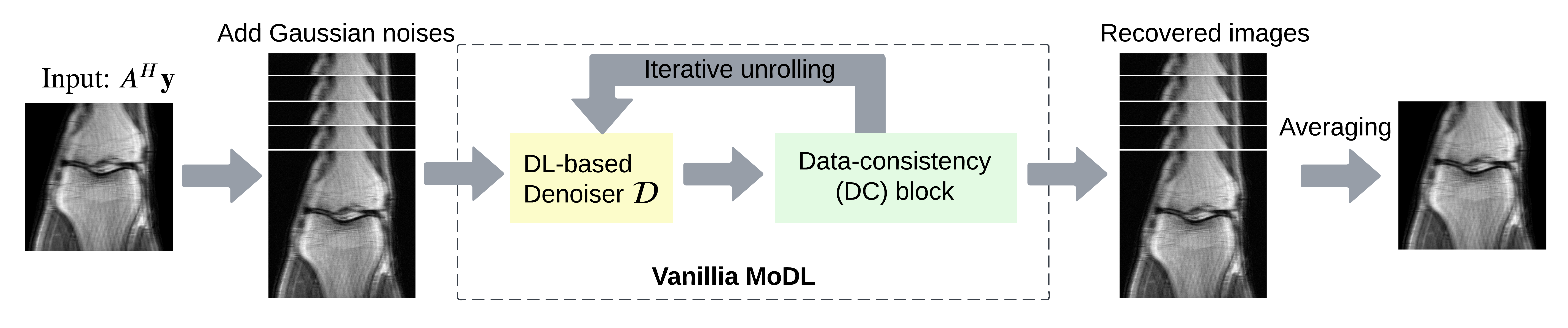}
        \caption{\footnotesize{A schematic overview of \ref{eq: denoised smoothing mri}.}}
        \label{fig: RS-E2E}
\end{figure}

\textbf{Fig.\,\ref{fig: RS-E2E}} provides an illustration of \ref{eq: denoised smoothing mri}-baked {\modl}.
Although \ref{eq: denoised smoothing mri} renders a simple application of RS to {\modl}, it remains unclear if \ref{eq: denoised smoothing mri} is the most effective way to bake RS into {\modl}, considering the latter's    learning specialities,
\textit{e.g.}, the involved denoising step 
and the DC step.
In the rest of the paper, we will focus on studying two main questions \textbf{(Q1)}--\textbf{(Q2)}.
 \begin{tcolorbox}[left=1.2pt,right=1.2pt,top=1.2pt,bottom=1.2pt]
\textbf{(Q1)}: \textit{Where should the RS operator   be integrated into {\modl}?}
\textbf{(Q2)}: \textit{How to design the denosier $\mathcal D(\boldsymbol \theta; \cdot)$   in the presence of RS?}
\end{tcolorbox}


\section{{\us}: SMoothed UnrollinG}
\label{sec: approach}

In this section, we tackle the above problems \textbf{(Q1)}--\textbf{(Q2)} by taking the unrolling characteristics of {\modl}  into the design of a RS-based robust MRI reconstruction. The proposed  novel integration of RS with {\modl} is termed 
{\textsc{\underline{Sm}oothed \underline{U}nrollin\underline{g}}}
(\textbf{\us}).


\subsection{Solution to (\textbf{Q1}): RS at intermediate unrolled denoisers}
\label{sec: unrolling}

\begin{figure}[htb]
    \centering
\begin{tabular}{c}
    \includegraphics[width=.45\textwidth]{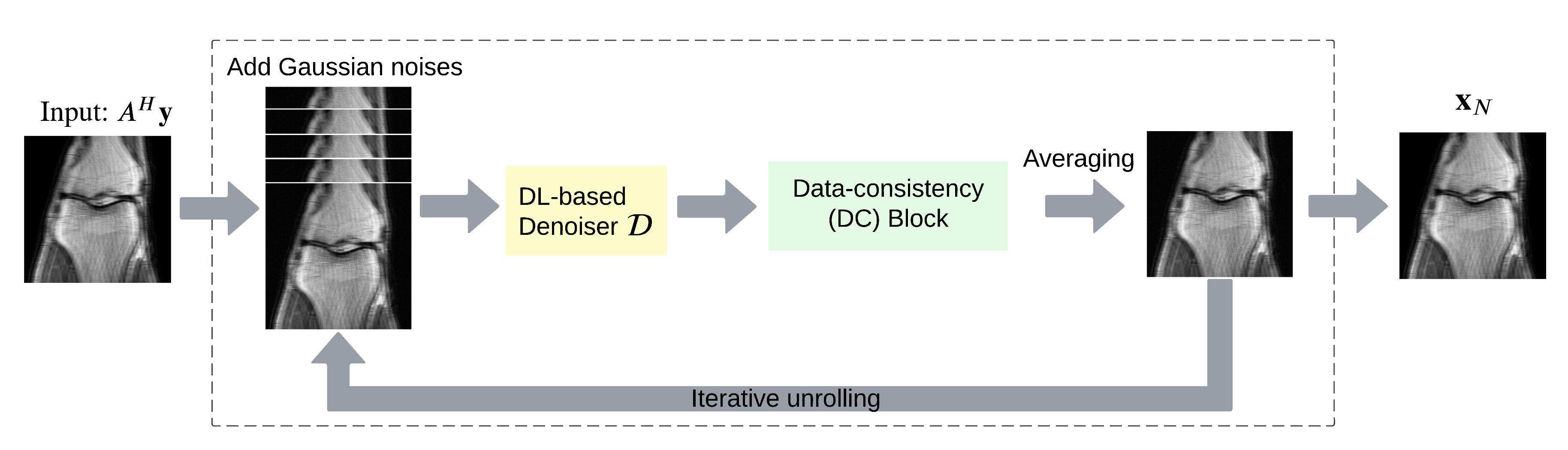}  \vspace*{-2mm}\\
   \footnotesize{(a) {\usold}  
        }\vspace*{2mm}\\
    \includegraphics[width=.45\textwidth]{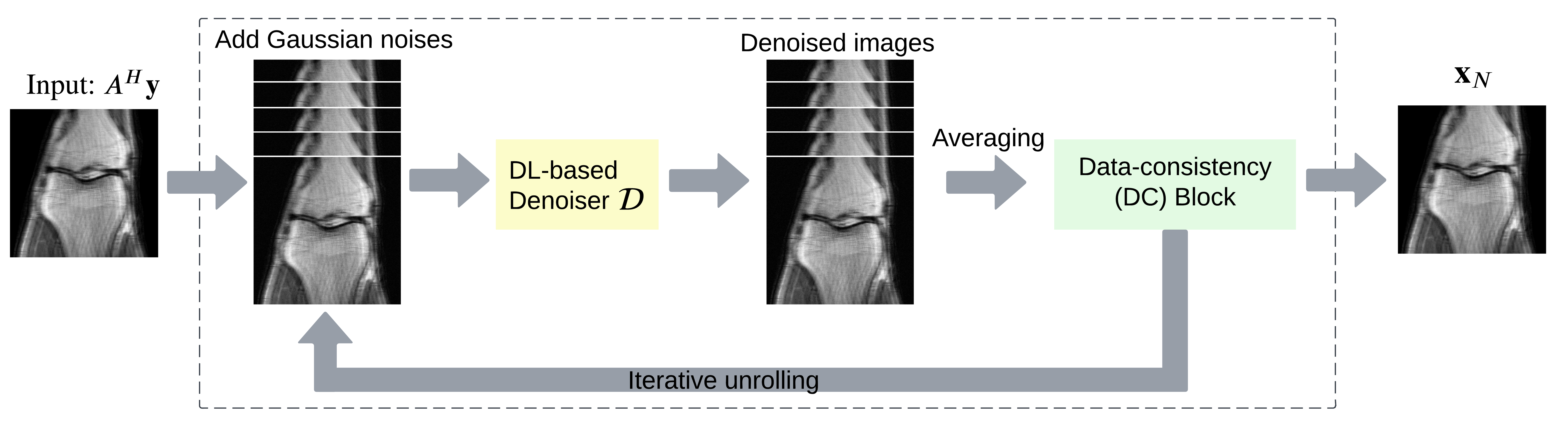} \vspace*{-2mm}\\
    \footnotesize{(b) \us}
\end{tabular}

    \vspace*{-1em}
    \caption{\footnotesize{Architectures of smoothed unrolling for {\modl}. 
    }}
    \label{fig: model-arch}
\end{figure}

Recall from \textbf{Fig.\,\ref{fig: RS-E2E}} that the RS operation is applied to {\modl} in an end-to-end fashion. Yet, the vanilla {\modl} framework  consists of multiple unrolling steps, each of which is naturally dissected into a \ding{172} \textit{denoising block} (denoted by $\mathcal D$)  and a \ding{173} \textit{DC block} (denoted by $\mathcal D \mathcal C$).  Taking the above architecture into account,  {RS} can also be integrated with each intermediate unrolling step of {\modl} instead of  following {\ref{eq: denoised smoothing mri}}. This leads to two \textit{new}   smoothing architectures of {\modl}:

\noindent \textbf{(a)  {\usold}}: In this scheme, the RS operation is incorporated into  {\modl} at each unrolled step (\textit{i.e.}, $\mathrm{RS}(\mathcal D + \mathcal{DC}) $). Formally, at the $n$th step, we have
$\mathrm{RS}(\mathcal D + \mathcal{DC}) =\mathbb{E}_{\boldsymbol \nu \sim\mathcal{N}(\mathbf 0, \sigma^2\mathbf I)} [ {\mathbf{x}}_{n} (\mathbf x_{n-1} + \boldsymbol \nu) ]  $, where 
${\mathbf{x}}_{n} (\mathbf x_{n-1} + \boldsymbol \nu)$ denotes the output of the $n$th unrolling step given the input $\mathbf x_{n-1}$ with Gaussian random noise $\boldsymbol \nu$. \textbf{Fig.\,\ref{fig: model-arch}-(a)} provides a schematic overview of {\usold}.

\noindent \textbf{(b)  {\us}}: Different from {\usold},  {\us} only applies RS  to the denoising network, leading to $\mathrm{RS}(\mathcal D) $ at each unrolling step. However, this seemingly simple modification aligns with  a robustness certification technique, called `denoised smoothing' \cite{salman2020denoised}, where a smoothed denoiser prepended to a victim model is sufficient to achieve   provable     robustness for this model. Formally, at the $n$th unrolling step, we have
\begin{align}
    \mathrm{RS}(\mathcal D)  =  \mathbb{E}_{\boldsymbol \nu \sim\mathcal{N}(\mathbf 0, \sigma^2\mathbf I)} [ \mathcal D_{\btheta} ( \mathbf x_{n-1} + \boldsymbol \nu) ] \Def \mathbf{z}_n,
    \label{eq: SMUG_RS}
\end{align}
together with the standard DC step 
$
\mathbf x_{n+1} = \argmin_{\mathbf x} 
  \|\mathbf A \bx - \by \|^{2}_2 + \lambda \| \bx - \mathbf z_n \|_2^2
$. \textbf{Fig.\,\ref{fig: model-arch}-(b)} shows the architecture of {\us}.

As will be evident later, our empirical results in Sec.\,\ref{sec: experiment} (\textit{e.g.}, \textbf{Fig.\,\ref{fig:archi_PSNR}}) show that {\us} and {\usold} can significantly outperform \ref{eq: denoised smoothing mri} in   adversarial robustness. In particular,  {\us} achieves the best robust performance without sacrificing its  standard  accuracy when evaluated on benign testing data. 


\subsection{Solution to \textbf{(Q2)}: {\us}'s pre-training and fine-tuning}
\label{sec: how to smoothing}
In what follows, we develop the training scheme of  {\us}. Spurred by the currently celebrated `pre-training + fine-tuning'  technique \cite{zoph2020rethinking,salman2020denoised}, we propose to train the {\us} model 
following this learning paradigm. 
Our rationale is that pre-training  is able to provide a  robustness-aware initialization of the DL-based denoising network for ease of fine-tuning.
{To pre-train the denoising network $\mathcal D_{\btheta}$, we consider a mean squared error (MSE) loss that 
measures the Euclidean distance between  images denoised by $\mathcal D_{\btheta}$ and the labels (\textit{i.e.}, target images, denoted by $\mathbf t$). This leads to the \textbf{pre-training} step:
\begin{equation}
   \btheta_\mathrm{pre} = \displaystyle \argmin_{\btheta} \mathbb{E}_{\mathbf t \in \mathcal D} [ \mathbb{E}_{\boldsymbol \nu }  || \mathcal D_{\boldsymbol \theta} (  \mathbf t + \nu)  -  \mathbf t ||_2^2]
\label{eq: pre-train_loss}
\end{equation}
where
$\mathcal{D}$ denotes the set of labels, 
$\boldsymbol \nu \sim\mathcal{N}(0,\sigma^2I)$. Note that the MSE loss \eqref{eq: pre-train_loss} does not engage the entire unrolled network. Thus, the pre-training is computational inexpensive and time-efficient.}


We next develop the fine-tuning scheme to improve  $\btheta_\mathrm{pre}$ based on labeled MRI datasets, \textit{i.e.}, with access to target images (denoted by $\mathbf t$). 
Since RS in  {\us}  (\textbf{Fig.\,\ref{fig: model-arch}-(b)})  is applied to every unrolling step, we propose an \textit{unrolled stability (\textbf{UStab}) loss} for fine-tuning the denoiser $\mathcal{D}_{\btheta}$:
\begin{align}
    \ell_{\mathrm{UStab}}(\btheta;   \mathbf y, \mathbf t)=
    \sum_{n=0}^{N-1} \mathbb{E}_{\boldsymbol \nu }||\cD_{\btheta}( \bx_n+\boldsymbol \nu)-\cD_{\btheta}(\mathbf t)||^2_2,
    \label{eq: unrolling loss}
\end{align}
where $N$ is the total number of unrolling steps, 
$\mathbf x_0 = \mathbf A^H \mathbf y $, and $\boldsymbol \nu \sim\mathcal{N}(0,\sigma^2I)$. {The UStab loss \eqref{eq: unrolling loss} relies on  target images, bringing in a key benefit: the denoising stability 
is guided by the  reconstruction accuracy of  the ground-truth image, yielding a graceful tradeoff between  robustness and accuracy. }

Integrating the UStab loss \eqref{eq: unrolling loss} with the vanilla reconstruction loss of {\modl} \cite{Aggarwal2019MoDL:Problems}, we obtain the \textbf{fine-tuned} $\btheta$ 
by using
\begin{equation}
   \displaystyle 
   \ell(\btheta; \mathbf y, \mathbf t)=\lambda_\ell
   \| 
   \mathbf x_N
   (\boldsymbol \theta; \mathbf A^H \mathbf y) - \mathbf t \|_2^2  +  \ell_{\mathrm{UStab}}(\btheta;   \mathbf y, \mathbf t) 
   , 
\label{eq: finetune_loss}
\end{equation}
where $\mathcal D$ denotes the labeled dataset, 
$ \mathbf x_N$
is the reconstructed image using {RS-applied} {\modl} ({\textit{i.e.}, {\usold} and {\us}}) with the denoising network of parameters $\btheta$ and input  $\mathbf A^H \mathbf y$,
and $\lambda_\ell > 0$ 
is a regularization parameter to strike the balance between  reconstruction error (for accuracy) and denoising stability (for robustness). 
We fine-tune $\btheta$
using $\btheta_\mathrm{pre}$ as   initialization.
\section{Experiments}
\label{sec: experiment}


\subsection{Experiment setup} 
\noindent \textbf{Models \& datasets.} 
The studied RS-baked {\modl} architectures are shown 
in \textbf{Figs.\,\ref{fig: RS-E2E}} and \textbf{\ref{fig: model-arch}}.
In experiments, we set  the total number of unrolling steps to $N = 8$, and set the denoising regularization parameter   $\lambda = 1$ in vanilla {\modl}. 
For the denoising network $\mathcal{D}_{\btheta}$, we  use the Deep Iterative Down-Up Network (DIDN) \cite{yu2019deep} with three down-up blocks and 64 channels. We adopt 
the conjugate gradient method \cite{Aggarwal2019MoDL:Problems}  with tolerance $1e^{-6}$ to implement the DC block. 
We conduct our experiments on 
the \texttt{fastMRI} dataset \cite{zbontar2018fastmri}. 
The observed data $\by$ are obtained with $15$ coils and are  cropped to the resolution of $320\times320$ for MRI reconstruction. To implement the observation model, we adopt a Cartesian mask  at $4\times$ acceleration (\textit{i.e.}, $25\%$ sampling rate). 
The coil sensitivity maps for all cases were obtained using the BART toolbox \cite{tamir2016generalized}.

\vspace*{1mm}
\noindent \textbf{Training \& evaluation.}
We use 304 images for training, 32 images for validation, and 64 images for testing (that are unseen during training). 
At \textbf{training time}, the batch size is set to $2$   trained on   two GPUs. 
We use the the Adam optimizer to train studied MRI reconstruction models with the momentum parameters $(0.5, 0.999)$.
The number of epochs is set to $60$ with a linearly decaying learning rate from $10^{-4}$ to $0$ after epoch 20. The stability parameter $\lambda_\ell$ 
in \eqref{eq: finetune_loss} is tuned so that  the standard accuracy  of the learned model is comparable to the vanilla {\modl}.
In RS, we set the standard deviation of Gaussian noise as 
$\sigma = 0.01$, and use $10$ Monte Carlo samplings to implement the smoothing operation. 
At \textbf{testing time}, 
we evaluate our methods on clean data, random noise-injected data and adversarial examples generated by 10-step PGD attack \cite{antun2020instabilities} of $\ell_\infty$-norm radius $\epsilon = 0.004$. The quality of reconstructed images is measured using  
peak signal-to-noise ratio (PSNR) and structure similarity (SSIM).
In addition to adversarial robustness, we also
evaluate the performance of our methods at the presence of another two perturbation sources (\textit{i.e.}, altered sampling rate and unrolling step number at testing time), as shown in \textbf{Fig.\,\ref{fig: robustness}}. 
 
 \begin{table}[htb]
\centering
\vspace*{-3mm}
\caption{\footnotesize{
Accuracy performance  of different smoothing architectures (\ref{eq: denoised smoothing mri}, {\usold}, {\us}), together with the vanilla  {\modl}. Here `Clean Accuracy', `Noise Accuracy', and `Robust Accuracy' refer to PSNR/SSIM evaluated on benign data, random noise-injected data, and PGD attack-enabled adversarial data, respectively.
$\uparrow$ signifies that a higher   number indicates a better reconstruction accuracy. The result $a${\tiny{$\pm b$}} represents mean $a$ and standard deviation $b$ over {64} testing images.
The relative performance is reported with respect to that of vanilla {\modl}.
}}
\label{tab: exp_smoothing}
\vspace*{-2mm}
\resizebox{0.48\textwidth}{!}{%
\begin{tabular}{c|cc|cc|cc}
\toprule[1pt]
\midrule
Models 
& \multicolumn{2}{c}{Clean Accuracy} 
& \multicolumn{2}{c}{Noise Accuracy} 
& \multicolumn{2}{c}{Robust Accuracy} 
\\
Metrics 
& PSNR \textcolor{red}{$\uparrow$} & SSIM \textcolor{red}{$\uparrow$}
& PSNR \textcolor{red}{$\uparrow$} & SSIM \textcolor{red}{$\uparrow$}
& PSNR \textcolor{red}{$\uparrow$} & SSIM \textcolor{red}{$\uparrow$}
\\
\midrule
Vanilla {\modl}
& 29.73\footnotesize{$\pm$3.27}
& 0.900\footnotesize{$\pm$0.07}
& 28.70\footnotesize{$\pm$2.77}
& 0.874\footnotesize{$\pm$0.07}
& 22.91\footnotesize{$\pm$2.42}
& 0.729\footnotesize{$\pm$0.07}
\\
\midrule
RS-E2E
& \textbf{+0.09}\footnotesize{$\pm$3.24}
& \textbf{+0.002}\footnotesize{$\pm$0.07}
& +0.38\footnotesize{$\pm$2.90}
& +0.010\footnotesize{$\pm$0.07}
& +0.78\footnotesize{$\pm$2.70}
& \textbf{+0.034}\footnotesize{$\pm$0.08}
\\
{\usold}  
& -1.01\footnotesize{$\pm$3.07}
& -0.014\footnotesize{$\pm$0.08}
& -0.09\footnotesize{$\pm$2.99}
& +0.008\footnotesize{$\pm$0.08}
& +3.08\footnotesize{$\pm$2.42}
& -0.014\footnotesize{$\pm$0.11}
\\
\rowcolor[gray]{.8}
{\us} (ours)
& -0.34\footnotesize{$\pm$3.06}
& -0.006\footnotesize{$\pm$0.08}
& \textbf{+0.53}\footnotesize{$\pm$2.98}
& \textbf{+0.016}\footnotesize{$\pm$0.08}
& \textbf{+3.87}\footnotesize{$\pm$2.28}
& +0.008\footnotesize{$\pm$0.11}
\\
\midrule
\bottomrule[1pt]
\end{tabular}%
}
\vspace*{-4mm}
\end{table}

\subsection{Experiment results}
   
   


\begin{wrapfigure}{r}{43mm}
\vspace*{-5mm}
    \centering
  \hspace*{-3mm}  \includegraphics[width=0.25\textwidth]{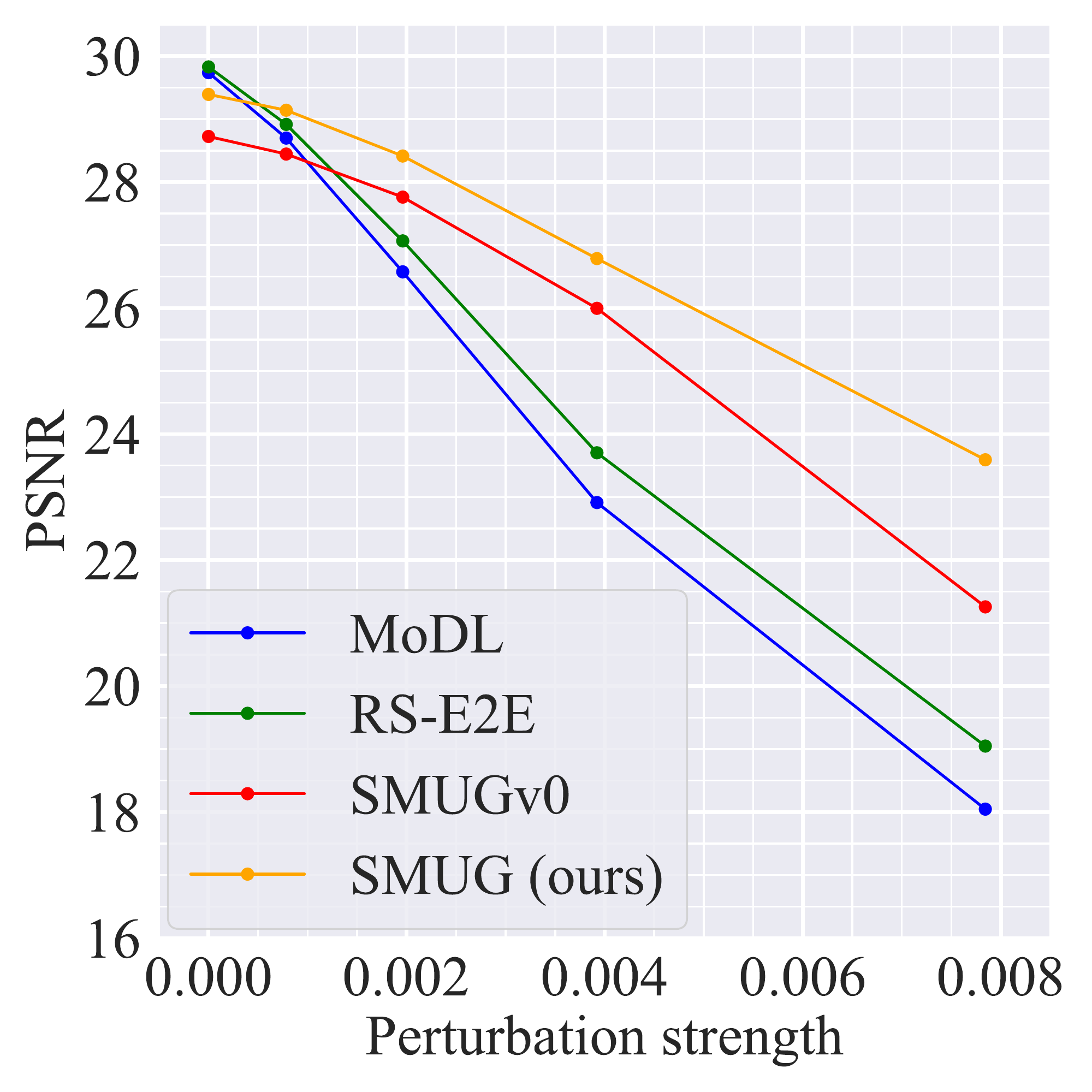}
    \vspace*{-3mm}
    \caption{\footnotesize{PSNR  of   baseline methods and proposed {\us} versus perturbation strength $\epsilon$ used in   PGD attack-generated adversarial examples at testing time. The case of $\epsilon =0$ corresponds to clean accuracy. 
    }}
    \label{fig:archi_PSNR}
  \vspace*{-5mm}
  \end{wrapfigure}
\begin{figure}[htb]

\begin{tabular}[b]{cccc}
        \includegraphics[width=.21\linewidth, trim=70 10 70 10]{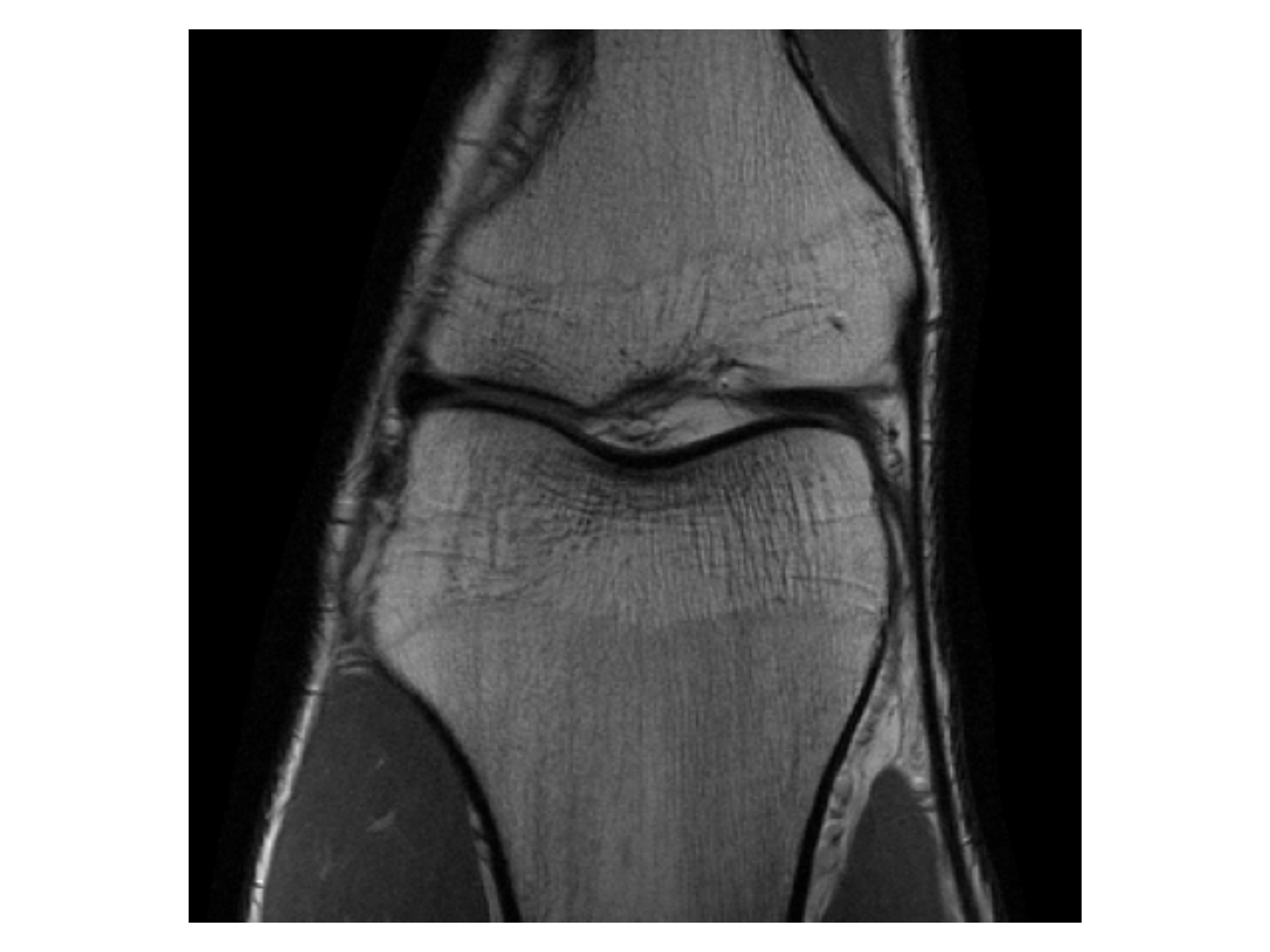}
        &\hspace{-0.2cm}
        \includegraphics[width=.21\linewidth, trim=70 10 70 10]{Figures/visualization/vanilla_MoDL_eps0.5_255.pdf}
        &\hspace{-0.2cm}
        \includegraphics[width=.21\linewidth, trim=70 10 70 10]{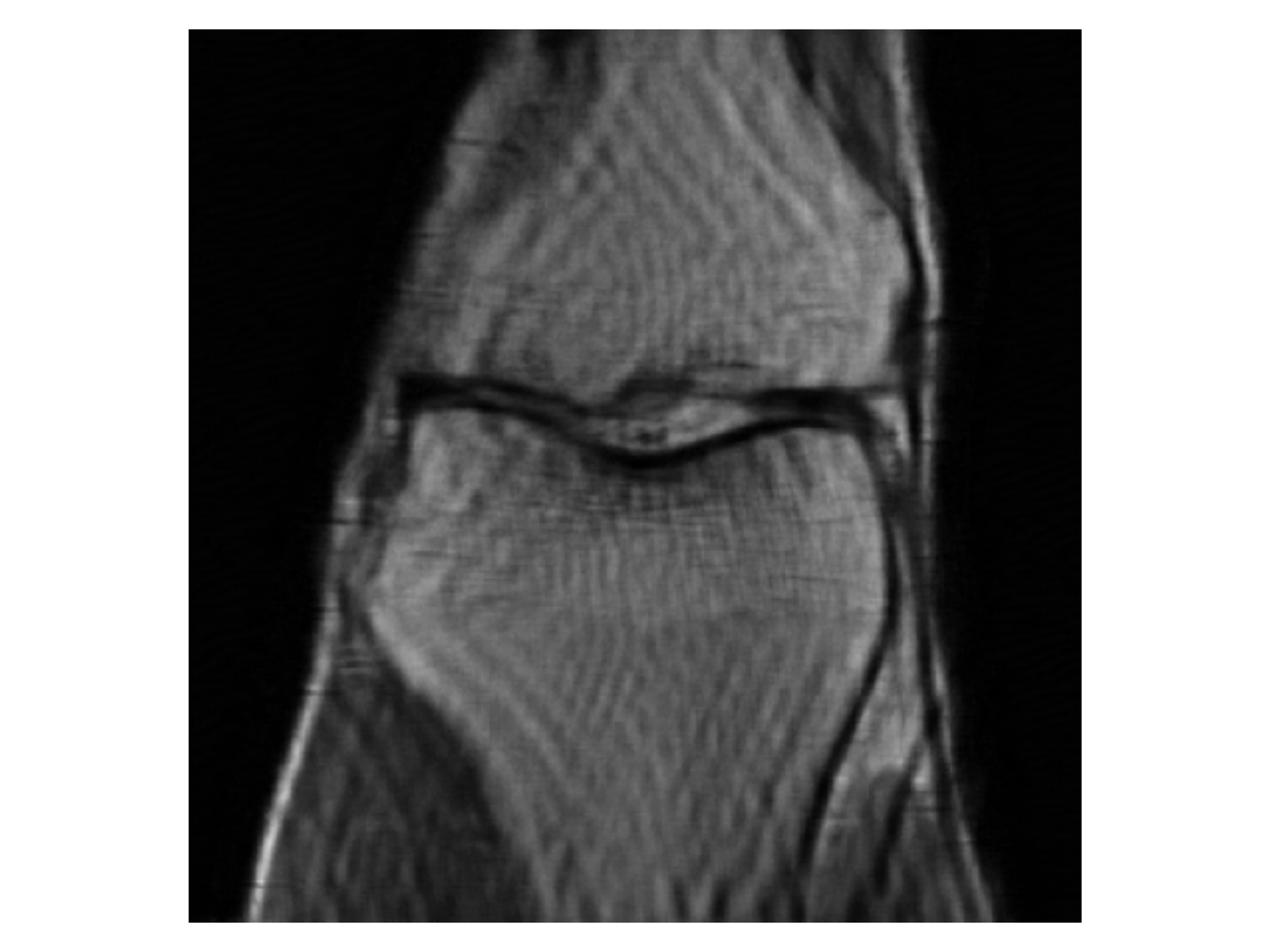}

        &\hspace{-0.2cm}
        \includegraphics[width=.21\linewidth, trim=70 10 70 10]{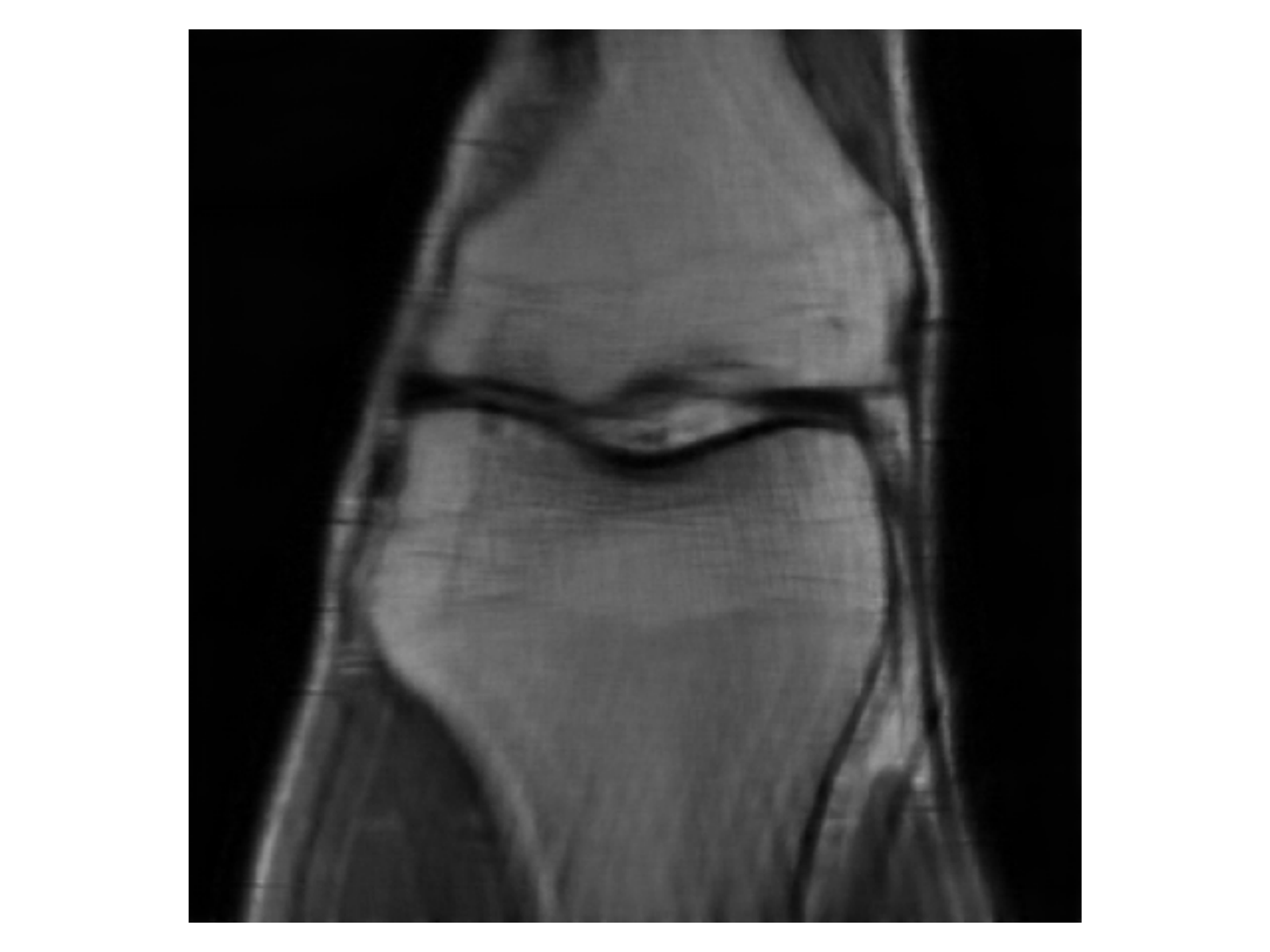}
        \\[-0pt]
        \scriptsize{(a) Ground Truth} 
        &\hspace{-0.2cm} 
        \scriptsize{(b) Vanilla {\modl}} 
        &\hspace{-0.2cm} 
        \scriptsize{(c) RS-E2E}
        &\hspace{-0.2cm} 
        \scriptsize{(d) \textsc{SMUG}} 
        \\ 
\end{tabular}
\vspace*{-4mm}
\caption{\footnotesize{Visualization of   ground-truth   and reconstructed images using different methods, evaluated on PGD attack-generated adversarial inputs of perturbation strength $\epsilon = 0.002$.
}}
\label{fig: vis}
\vspace*{-2mm}
\end{figure}

\textbf{Table\,\ref{tab: exp_smoothing}} shows PSNR and SSIM values for different smoothing architectures with different training schemes, along with vanilla {\modl} as a baseline, evaluated on clean and adversarial test datasets. We present the PSNR results for these models under different scales of adversarial perturbations (\textit{i.e.}, attack strength $\epsilon$) in \textbf{Fig.\,\ref{fig:archi_PSNR}}.
We observe that our method {\us} outperforms all other models in robustness, consistent with the visualization of reconstructed images in \textbf{Fig.\,\ref{fig: vis}}. Also, {\us} yields a promising clean accuracy performance, which is better  than {\usold} and   comparable to the vanilla {\modl} model. This shows the effectiveness of our proposed method for improving robustness while preserving clean accuracy \emph{(i.e., without the perturbations)}.   

\begin{figure}[htb]
\vspace*{-3mm}
    \centering
\begin{tabular}{cc}
  \hspace*{-2mm}  \includegraphics[width=0.24\textwidth]{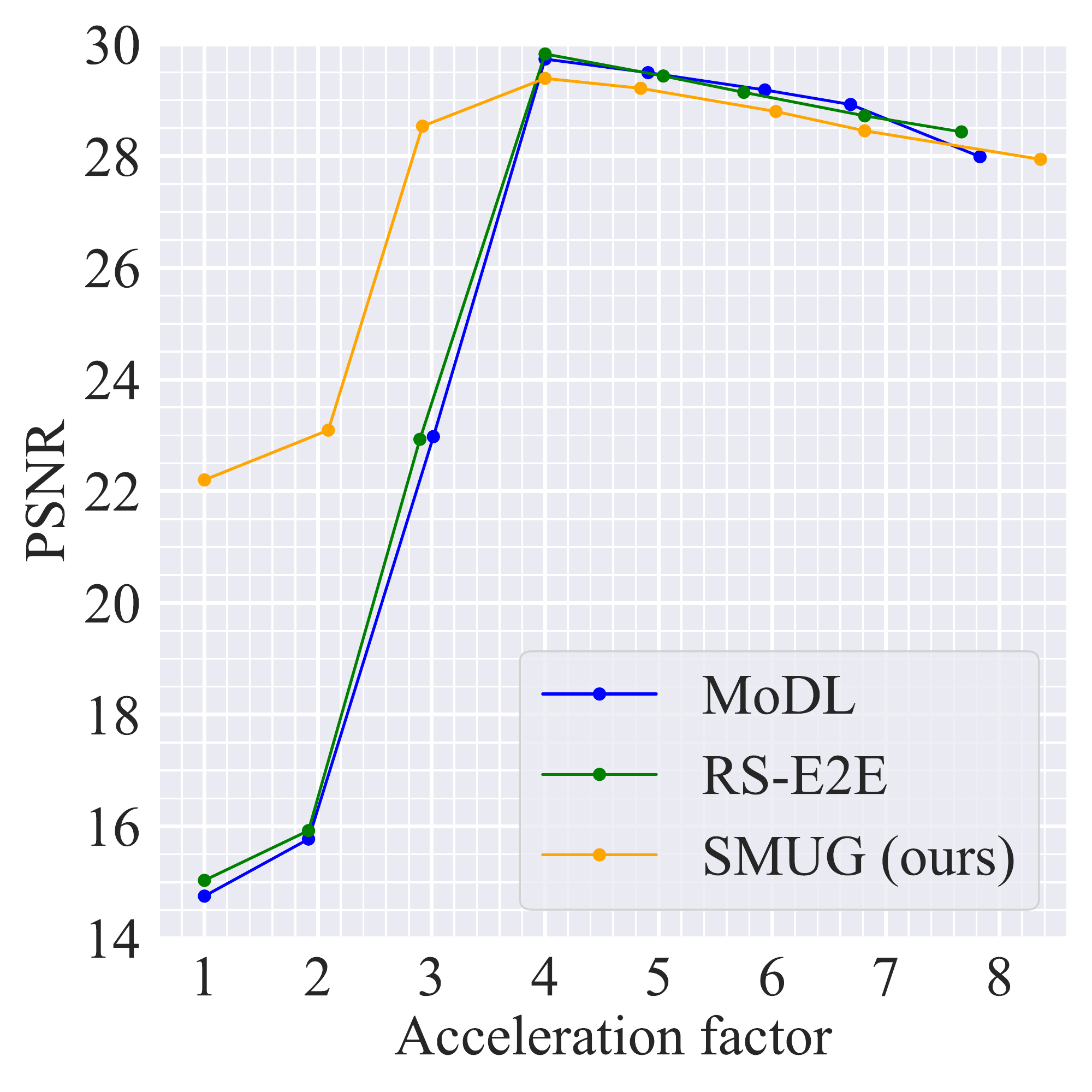}   &    \hspace*{-5mm} \includegraphics[width=0.24\textwidth]{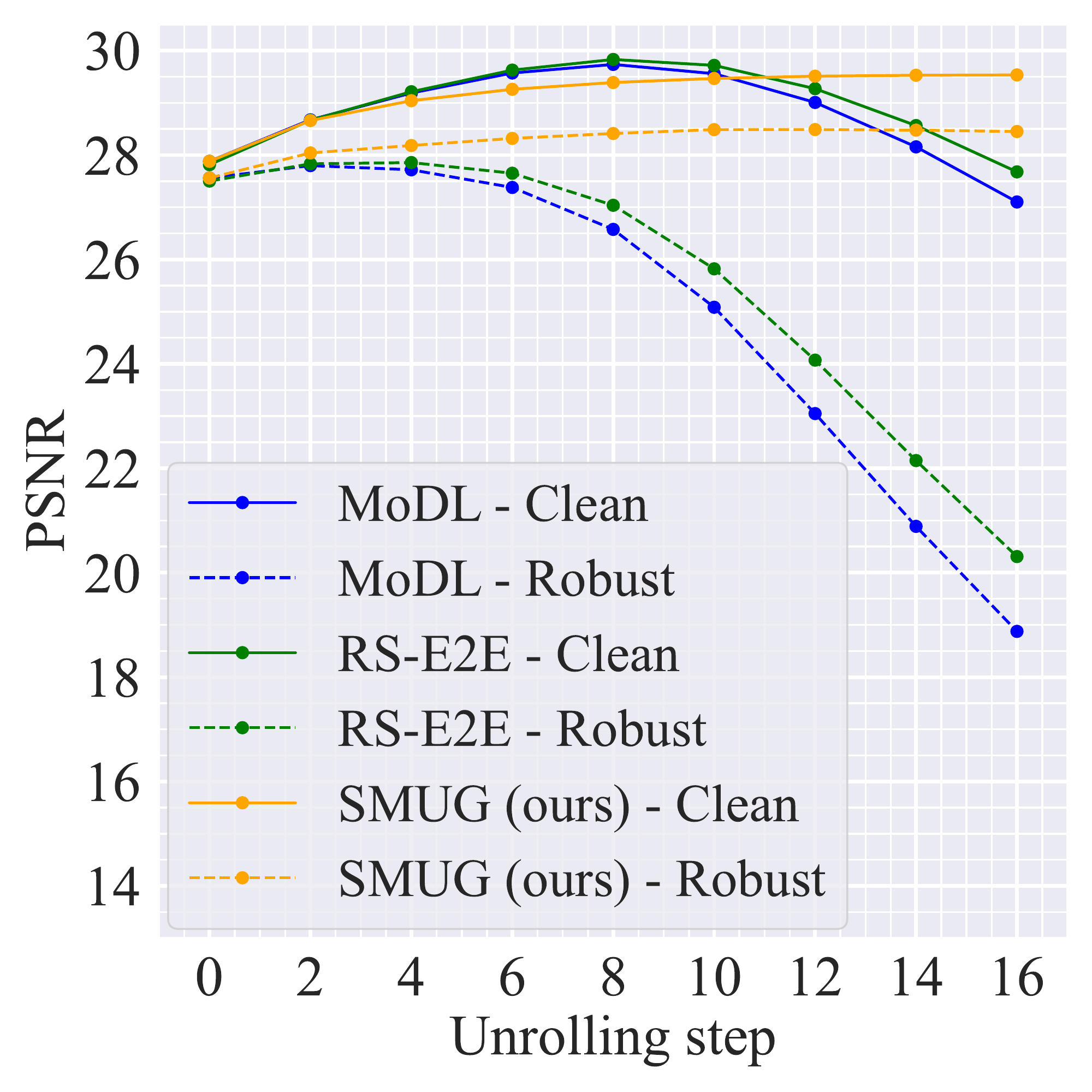}
\end{tabular}
    \vspace*{-5mm}
    \caption{\footnotesize{
    PSNR results of different MRI reconstruction methods versus
    different measurement sampling rates ($4\times$ acceleration \textit{i.e.}, $25\%$ sampling rate at training; Left plot) and
    unrolling steps (8 at training; Right plot). 
    }}
    \label{fig: robustness}
    \vspace*{-2mm}
\end{figure}

Next, we evaluate the  effectiveness  of MRI reconstruction methods when facing sampling rate and unrolling step perturbations at testing time.  In other words, there exists a test-time shift for the training  setup of MRI reconstruction.  
In \textbf{Fig.\,\ref{fig: robustness}}, we present the evaluation results of {\us}, with two baselines, vanilla {\modl} and \ref{eq: denoised smoothing mri}, on different unrolling steps and sampling rates. Note that   these models are trained with the number of unrolling steps $K=8$ and sampling masks with the $4\times$ {acceleration (\textit{i.e.}, 25\% sampling rate).}
As we can see, {\us} achieves a remarkable improvement in robustness against different sampling rates and unrolling steps, which {\modl} and \ref{eq: denoised smoothing mri} fail to achieve.
Although we do not intentionally design our method to mitigate {\modl}'s instabilities against perturbed sampling rate and unrolling step number, {\us} still provides improved PSNRs over  other baselines. 
We credit the improvement to the close relationships between these two instabilities with adversarial robustness. 

\begin{wrapfigure}{r}{43mm}
\vspace*{-6mm}
\centering
  \hspace*{-3mm}  \includegraphics[width=0.25\textwidth]{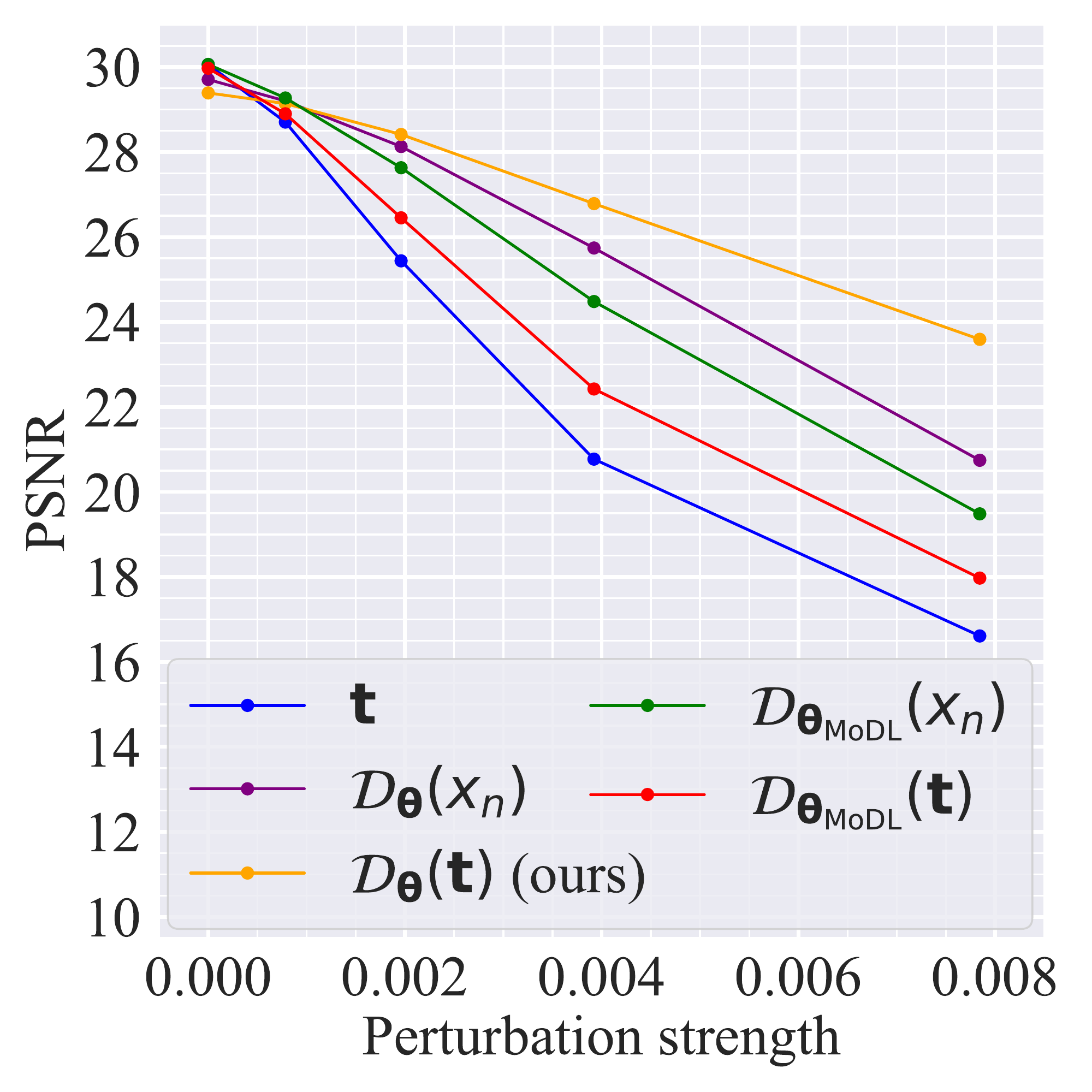}
    \vspace*{-4mm}
    \caption{\footnotesize{PSNR vs. adversarial attack strength ($\epsilon)$ of {\us} for different configurations of   UStab loss \eqref{eq: unrolling loss}. 
    }}
    \label{fig: reference_PSNR}
  \vspace*{-5mm}
\end{wrapfigure}
We conduct additional experiments showing the importance of integrating    target image denoising into {\us}'s training pipeline in \eqref{eq: unrolling loss}. \textbf{Fig.\,\ref{fig: reference_PSNR}} shows PSNR versus perturbation strength ($\epsilon$) when using different alternatives to $\mathcal D_{\btheta} (\mathbf t)$ in~\eqref{eq: unrolling loss}, including 
$\mathbf t$ (the original target image), $\mathcal D_{\btheta}(\mathbf x_n)$ (denoised   output of each unrolling step), and their variants when using the fixed, vanilla {\modl}'s denoiser $\mathcal D_{\btheta_\text{\modl}}$ instead.
As we can see, the performance of {\us} varies when the UStab loss \eqref{eq: unrolling loss} is configured differently. The proposed $\mathcal D_{\btheta}(\mathbf t)$ outperforms the other baselines. A possible reason is that it infuses supervision of target images in an adaptive, denoising-friendly manner, \textit{i.e.}, taking influence of $\mathcal D_{\btheta}$ into consideration.


\vspace*{-4mm}

\section{Conclusion}
In this work, we proposed a scheme for improving robustness of DL-based MRI reconstruction. We showed deep unrolled reconstruction's ({\modl}'s) weaknesses in robustness against adversarial perturbations, sampling rates, and unrolling steps. To improve the robustness of {\modl}, we proposed {\us} with a novel unrolled smoothing loss.
 Compared to the vanilla
{\modl} approach and several variants of {\us}, we empirically showed that
our approach is effective and can   significantly improve the robustness of {\modl} against a diverse set of external perturbations. In the future, we will study the problem of certified robustness  and derive the certification bound of adversarial perturbations using  randomized smoothing.



\clearpage
\newpage 
\bibliographystyle{IEEEbib}
\bibliography{reference,refs_adv}

\end{document}